\documentclass[%
 preprint,
 superscriptaddress,
 amsmath,amssymb,
 aps,
 showkeys,
 titlepage,
 endfloats*.   %move all floats to the end of the document and put each on a separate page
]{revtex4-2}

\usepackage[utf8]{inputenc}
\usepackage[english]{babel}
\usepackage{longtable}
\usepackage{graphicx}% Include figure files
\usepackage{dcolumn}% Align table columns on decimal point
\usepackage{bm}% bold math

\usepackage[colorlinks = true,
            linkcolor = blue,
            urlcolor  = blue,
            citecolor = red,
            anchorcolor = blue]{hyperref}

\begin{document}

%\preprint{APS/123-QED}

\title{High-throughput Discovery of Anti-gap Semiconductors} 

\author{Zeyu Xiang}
\affiliation{Department of Mechanical Engineering, University of California, Santa Barbara, CA 93106, USA}

\author{Fanghao Zhang}
\affiliation{Department of Mechanical Engineering, University of California, Santa Barbara, CA 93106, USA}

\author{Bolin Liao}
\email{bliao@ucsb.edu} \affiliation{Department of Mechanical Engineering, University of California, Santa Barbara, CA 93106, USA}

\date{\today}

\begin{abstract}
Conventional semiconductors typically have bonding states near the valence band maximum (VBM) and antibonding states near the conduction band minimum (CBM). Semiconductors with the opposite electronic configuration, namely an antibonding VBM and a bonding CBM, are here termed ``anti-gap semiconductors". They have been theoretically proposed to exhibit excellent optoelectronic properties because of their strong tolerance to defects. However, no anti-gap semiconductors have been identified so far, despite a known list of semiconductors with an antibonding VBM. Here, we use high-throughput computation to identify over 100 anti-gap semiconductors. From this group, we analyze the transition metal dichalcogenide MX$_2$ (M=Hf, Zr; X=S, Se) family in detail. In addition to verifying their defect tolerance for both electrons and holes using first-principles simulations, we also discovered that photoexcitation of charge carriers can lead to significant lattice stiffening and increased thermal conductivity in anti-gap semiconductors, which can be potentially used as photo-driven thermal switches. Our work analyzes the formation of the anti-gap electronic structure and showcases their unusual photoinduced lattice dynamics that can have a potential impact on their photophysical applications.  
\end{abstract}

%Use showkeys class option if keyword display desired
                            
\maketitle

%\tableofcontents

% \section{Introduction}
% In general: please start a newline for each sentence: this helps locate accurate positions in the manuscript using Overleaf.

% Use this format for citation keys: first-author-last-name/year/first-word-of-title. This is the default format for google scholar import.

% Introduction is the most important part of a paper and should be a more detailed version of the abstract. An exemplar organization of the Introduction:

% \textbf{[Paragraph 1: General Background.] }

\begin{figure}[!htb]
\includegraphics[width=\textwidth]{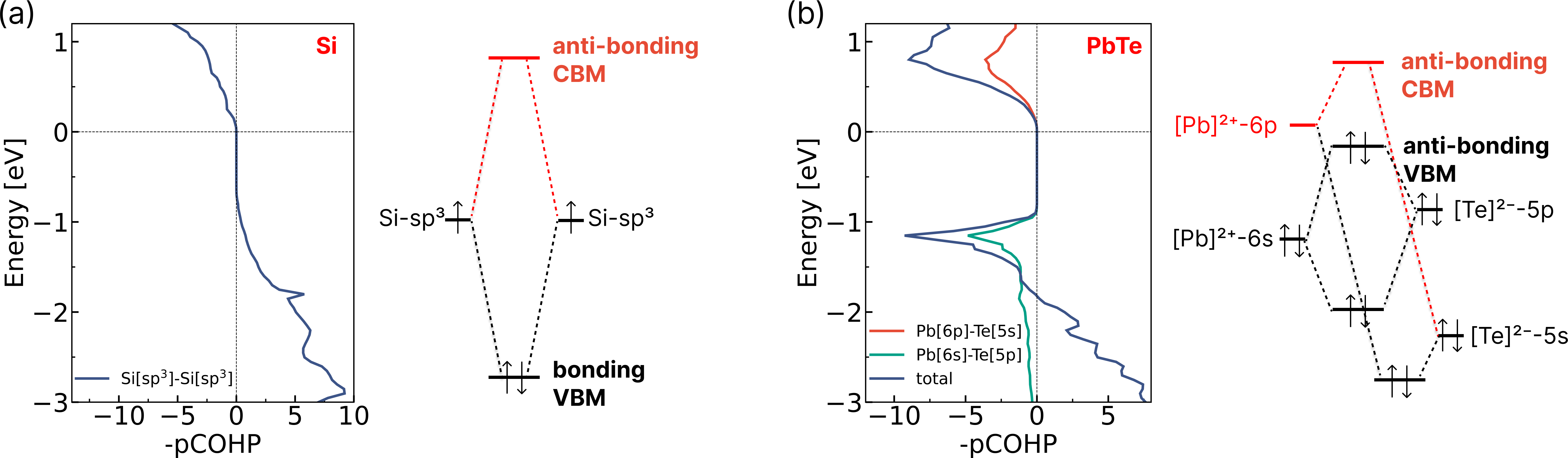}
\caption{\textbf{COHP analysis and schematic diagrams of orbital interactions for Si and PbTe.}
(a) COHP diagram of Si, showing the bonding VBM and antibonding CBM.
The orbital interaction diagram shows that sp$^3$ hybridization in Si forms $\sigma$ and $\sigma^*$ bonds (bonding VBM and antibonding CBM).
(b) COHP diagram of PbTe, showing the antibonding VBM and antibonding CBM.
The orbital interaction diagram shows that Pb-6s and Te-5p states hybridize to form the antibonding VBM and Pb-6p and Te-5s states hybridize to form the antibonding CBM.
} 
\label{fig:Si_PbTe}
\end{figure}

The macroscopic properties of materials are intimately related to the chemical bonding of their constituents at the microscopic level~\cite{burdett1995chemical}. In semiconductors, in particular, the intricate relationship between bonding types and physicochemical propertiesmanifests prominently in the bonding and antibonding states near the band edges, specifically at the valence band maximum (VBM) and conduction band minimum (CBM). Common semiconductors, such as Si and the III-V family, feature bonding states at VBM and antibonding states at CBM. For example, in Si, the bonding VBM and antibonding CBM result from sp$^3$ orbital hybridization forming $\sigma$ and $\sigma^*$ bonds, as illustrated in Fig.~\ref{fig:Si_PbTe}a. Strong covalent bonds formed this way lead to stiff phonons and high thermal conductivity. Significant orbital overlap and delocalization of electronic wavefunctions also contribute to their relatively high charge mobility. In these materials, photoexcitation of electrons from the VBM to the CBM reduces the occupancy of stable bonding states and increases the occupancy of unstable antibonding states. As a result, photoexcitation in these materials can lead to lattice destabilization and even phase transition, often accompanied by the softening of phonons. For example, nonthermal melting of semiconductors such as Si and GaAs due to intense photoexcitation has been observed experimentally~\cite{shank1983time,sundaram2002inducing,siders1999detection,rousse2001non} and attributed to photoinduced bond weakening and the associated phonon softening~\cite{stampfli1990theory,silvestrelli1996ab,recoules2006effect}. 

Semiconductors hosting an antibonding VBM have recently attracted intense research interests because of their unusual, and often desirable, properties. Partial oxidation of cations is a common mechanism to form an antibonding VBM. For example, partially oxidized Pb$^{2+}$ ions with a lone pair of 6s electrons found in IV-VI semiconductors (PbTe, PbSe, PbS)~\cite{waghmare2003first} and lead halide perovskites~\cite{fabini2020underappreciated} are known to create antibonding VBM states. In these materials, the occupied 6s lone pair states hybridize with occupied anion p states to form a filled anti-bonding valence band. Strong hybridization of the Pb 6p states with the anion s states, however, also leads to an empty antibonding conduction band. As an example, the bonding scheme in PbTe is illustrated in Fig.~\ref{fig:Si_PbTe}b. Another example is the formation of antibonding VBMs due to p-d hybridization in compounds containing Cu$^+$ or Ag$^+$ ions~\cite{zakutayev2014defect,he2022accelerated}. Occupied antibonding VBM states are shown to lead to soft and anharmonic phonons and low lattice thermal conductivity~\cite{yuan2023lattice,das2023strong,he2022accelerated}, which can benefit thermoelectric applications. In the meantime, the antibonding VBM states are believed to be linked to the extraordinary photovoltaic performance of lead halide perovskites, owing to factors including defect tolerance and a prolonged minority carrier lifetime~\cite{fabini2020underappreciated,liu2021antibonding}. Furthermore, photoexcitation reduces the occupancy of unstable antibonding VBM states and can lead to strong photostriction in these materials~\cite{xiang2024high}. So far, all of these materials with antibonding VBM states also host antibonding CBM states.

In this work, we examine the possibility of the opposite electronic configuration to that of common semiconductors, namely semiconductors with an antibonding VBM and a bonding CBM. We term these materials ``anti-gap semiconductors''.
Previously, these materials have been theoretically proposed to exhibit strong defect tolerance through the resonant alignment of defect states within the bands~\cite{zakutayev2014defect,brandt2015identifying,ganose2016relativistic,brandt2017searching,liu2021antibonding}, avoiding the formation of deep in-gap states. However, to the best of our knowledge, no material with such an electronic configuration has been identified so far (Zakutayev et al. proposed Cu$_3$N as such a material~\cite{zakutayev2014defect}, but detailed calculation reveals that Cu$_3$N hosts both antibonding VBM and CBM. See the Supplementary Material for more analysis). In addition to defect tolerance, we expect that anti-gap semiconductors also possess unusual photophysical properties: Photoexcitation in an anti-gap semiconductor moves electrons from unstable antibonding states at the VBM to stable bonding states at the CBM, leading to stiffening of the bonds and potentially a metastable state after lattice relaxation. This behavior is expected to influence both phonon transport and photocarrier recombination properties.  

Here, we leverage first-principles density functional theory (DFT) calculations with the crystal orbital Hamilton population (COHP) method~\cite{dronskowski1993crystal,deringer2011crystal} to search for anti-gap semiconductors. COHP analysis decomposes the energy of the electronic band structure into interactions (overlaps) between pairs of atomic orbitals between adjacent atoms, allowing for a quantitative measure of the bonding and anti-bonding contributions to the electronic bands, especially near the band edges.
By convention, a positive (negative) sign indicates anti-bonding (bonding) interactions and COHP diagrams plot the negative value (-pCOHP), therefore making bonding (anti-bonding) states on the right (left) of the axis for intuitive visualization~\cite{deringer2011crystal}. COHP diagrams computed for Si and PbTe are provided in Fig.~\ref{fig:Si_PbTe}a and ~\ref{fig:Si_PbTe}b.
Using COHP, we performed a high-throughput screening of 15,000 semiconductors (with a bandgap within 5 eV) in the Materials Project~\cite{jain2013commentary}, identifying 117 semiconductors without rare-earth elements that exhibit anti-gap characteristics. Rare-earth elements were excluded because of the known inaccuracy of DFT simulating them.
From this search, the van der Waals (vdW) layered transition metal dichalcogenide (TMD) family, MX$_2$ (M=Hf, Zr; X=S, Se), both in their bulk and monolayer form, was identified as a prominent class of candidates. A detailed analysis was then conducted on the formation of the anti-gap characteristic in these materials.
Additionally, we verified the defect tolerance of these materials computationally by showing that vacancies create shallow defect levels within the bands instead of deep in-gap defects. 
Finally, we show that these anti-gap semiconductors exhibit a photoinduced stiffening of phonons and significant enhancement in thermal conductivity, in contrast to photoinduced lattice destabilization in conventional semiconductors. Our result provides the first examples of anti-gap semiconductors and points out their several unusual properties with potential applications in optoelectronics and thermal management.

\begin{figure}[!htb]
\includegraphics[width=\textwidth]{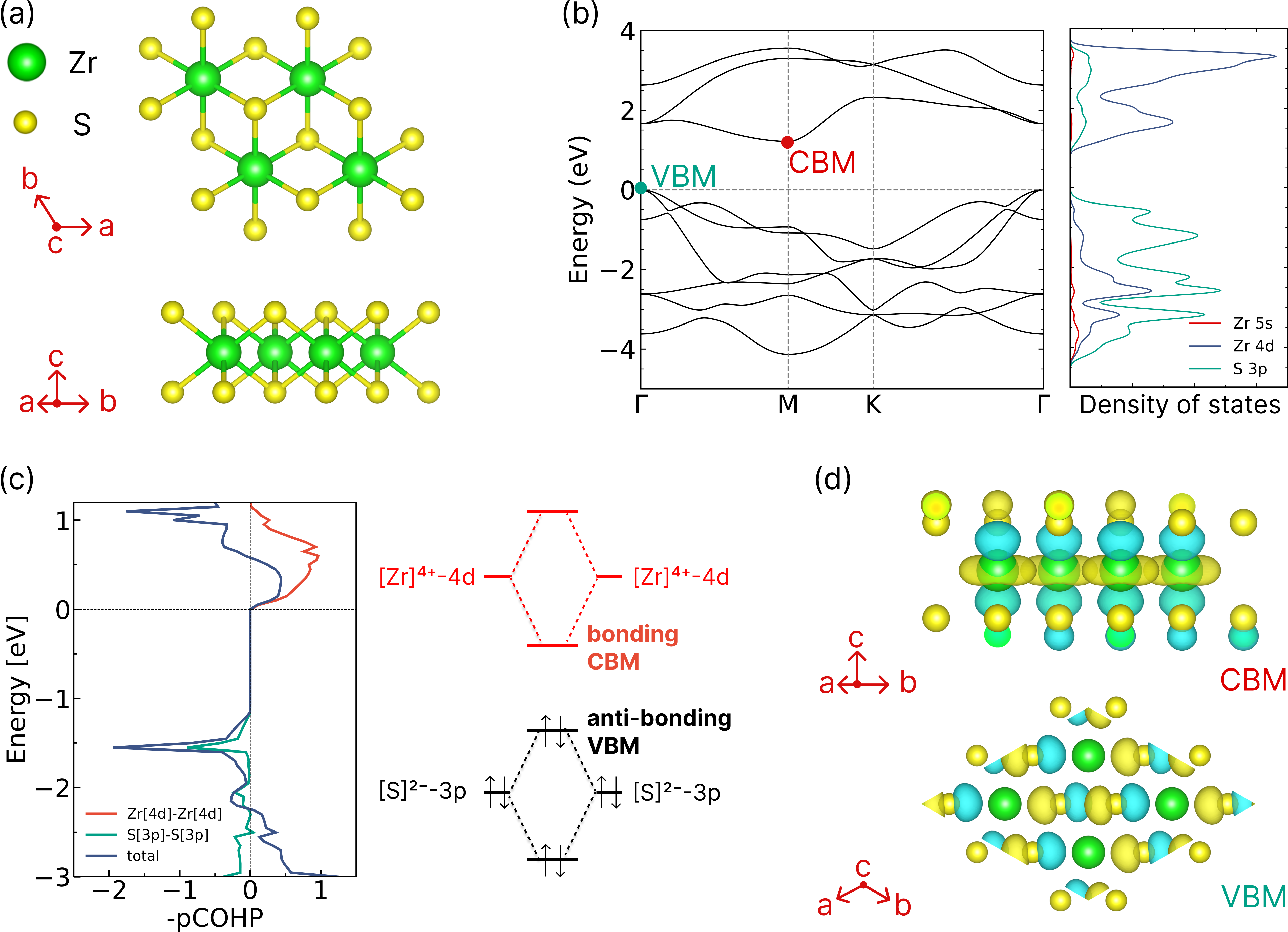}
\caption{\textbf{Structure and electronic properties of anti-gap semiconductor monolayer ZrS$_2$.}
(a) Top view and side view of the atomic structure of monolayer ZrS$_2$.
(b) Calculated electronic band structure and projected electronic density of states of monolayer ZrS$_2$.
(c) Isosurfaces of the electronic wavefunctions at VBM and CBM in monolayer ZrS$_2$.
The p orbitals between adjacent S atoms  exhibit antibonding interactions, as indicated by the opposite signs of their wavefunctions.
The d orbitals between adjacent Zr atoms exhibit bonding interactions, as evidenced by the same signs of their wavefunctions.
(d) COHP diagram of monolayer ZrS$_2$, showing the antibonding VBM and bonding CBM. 
The orbital interaction diagram shows that S-3p and S-3p states hybridize to form the antibonding VBM and Zr-4d and Zr-4d states hybridize to form the antibonding CBM.
} 
\label{fig:iso_ZrS2}
\end{figure}

Calculation details and parameters used in various steps are provided in the Supplementary Material. During the high-throughput material screening, we focus on the COHP character of the band edges of each material and aim to discover anti-gap semiconductors.
The discovered 117 anti-gap semiconductors are listed in the Supplementary Material. In this paper, we analyze the MX$_2$ (M=Hf, Zr; X=S, Se) family in detail due to their relatively simple structure and chemistry. In addition, both the bulk and the monolayer forms of this family possess the anti-gap electronic structure.
Taking ZrS$_2$ as an example, its bulk form has a layered trigonal CdI$_2$ type structure. Figure~\ref{fig:iso_ZrS2}a presents the top view and the side view for the crystal structure of monolayer ZrS$_2$, where Zr atoms are arranged in a triangular network, each coordinated with six S atoms forming edge-sharing ZrS$_6$ octahedra. Zr atoms sit at the center of the octahedra, which preserves the inversion symmetry, in contrast to the structure of MoS$_2$.
This configuration results in a triple-layered (S–Zr–S) structure, characteristic of the 1T-phase TMDs.
The monolayer lattice constant is 3.68 \AA\ after relaxation, same with previous theoretical calculations~\cite{zhuang2013computational}.
Figure~\ref{fig:iso_ZrS2}b shows the electronic band structure of monolayer ZrS$_2$ together with the electronic density of states (DOS) projected onto atomic orbitals.
For monolayer ZrS$_2$, the VBM is located at the $\Gamma$ point, while the CBM appears at the $\mathrm{M}$ point, resulting in an indirect bandgap of 1.19 eV.
This result is in good agreement with a previous theoretical calculation~\cite{li2014indirect}.
Experimental measurements of bandgaps for bulk ZrS$_2$ are 1.70 eV~\cite{moustafa2009growth} and 1.68 eV~\cite{greenaway1965preparation}.
The electronic DOS projected onto Zr-5s, Zr-4d, and S-3p orbitals reveals that the valence band is predominantly derived from the S-3p states, while the conduction band has main contributions from the Zr-4d states. The small projected DOS of Zr-4d and Zr-5s states in the valence band suggests a strongly ionic interaction between Zr and S, through which the Zr-5s and Zr-4d electrons are transferred to S-3p states. The corresponding COHP diagram of monolayer ZrS$_2$ is provided in Fig.~\ref{fig:iso_ZrS2}c, highlighting the importance of d-d interactions between Zr atoms and p-p interactions between S atoms in forming the anti-gap feature. A schematic of the bonding scheme is shown accordingly in Fig.~\ref{fig:iso_ZrS2}c. 
This bonding scheme is in contrast to the archetypal TMD MoS$_2$, whose COHP diagram and bonding scheme are shown in Fig. S3, where interactions between partially occupied Mo-4d states lead to a bonding valence band while strong hybridization between Mo-4d and S-3p gives rise to anti-bonding CBM states. To directly visualize the anti-bonding/bonding nature of the VBM/CBM states in monolayer ZrS$_2$, Fig.~\ref{fig:iso_ZrS2}d shows the wavefunctions at VBM and CBM, respectively, where the yellow and blue isosurfaces indicate opposite signs of the wavefunctions.
The opposite signs of the isosurfaces between p orbitals from adjacent S atoms confirm the anti-bonding nature of VBM, whereas the same sign of the isosurfaces between d orbitals of adjacent Zr atoms signifies a bonding character in CBM. We note here that the dominant orbital interactions in monolayer ZrS$_2$ are sufficiently strong to cause significant electronic band dispersion and small effective masses (listed in Table SI and Table SII in the Supplementary Material), confirming its semiconducting nature. Similar results of both bulk and monolayer forms of the other members of the family are provided in the Supplementary Material.

From the preceding discussion, we propose the following conditions that can potentially lead to anti-gap configurations: (1) relatively strong ionic interactions between cations and anions that facilitate a complete valence electron transfer from the cations to the anions; (2) strong covalent interactions among empty cation states to form a bonding CBM and occupied anion states to form an anti-bonding VBM. In Table~\ref{tab:sumrule}, we list several transition metal chalcogenides with anti-gap configurations that satisfy (1), which notably include known materials with promising photovoltaic properties such as BaZrS$_3$~\cite{agarwal2025synthesis}. To satisfy (2), the distance between neighboring cations and anions should be sufficiently small to ensure significant orbital overlap, which scales with $d^{-2}$ with $d$ being the interatomic distance. However, if the interatomic distance becomes too small, the hybridization energy shifts may become larger than the energy difference between isolated cation and anion orbitals that form the band edges. For example, the Mo-Mo distance (3.2\AA) in MoS$_2$ is shorter than the Zr-Zr distance (3.7\AA) in ZrS$_2$, resulting in deeper stabilization of the bonding d–d orbitals and leads to the formation of bonding valence states in MoS$_2$. The same applies to the distance between anions. In Table~\ref{tab:sumrule}, we observe that the shortest distances between anions in the group of anti-gap transition metal chalcogenides fall in the range of 3.5 to 3.7 \AA. We caution that these guidelines stem from an empirical analysis of our screening results, and there may be other possible scenarios to form anti-gap configurations.

\begin{table}[h] 
    \centering 
    \caption{Examples of transition metal chalcogenides with anti-gap configuration}
    \label{tab:sumrule}  
    \begin{tabular}{p{4cm}p{6cm}c}
        \hline
        Material & Cation electronic configurations & Shortest Anion Distance (\AA) \\  % Table header
        \hline
        ZrTl$_2$S$_3$ & Zr(4d$^2$5s$^2$) Tl(6p$^1$) & 3.52\\
        Hf$_3$Tl$_2$(CuSe$_4$)$_2$ & Hf(5d$^2$6s$^2$) Tl(6p$^1$) Cu(4s$^1$) & 3.68\\
        Zr$_3$Tl$_2$(CuS$_4$)$_2$ & Zr(4d$^2$5s$^2$) Tl(6p$^1$) Cu(4s$^1$) & 3.53\\
        Zr$_3$Tl$_2$(CuSe$_4$)$_2$ & Zr(4d$^2$5s$^2$) Tl(6p$^1$) Cu(4s$^1$) & 3.72\\
        NaHfCuSe$_3$ & Na(3s$^1$) Hf(5d$^2$6s$^2$) Cu(4s$^1$) & 3.63\\
        HfTl$_2$Se$_3$ & Hf(5d$^2$6s$^2$) Tl(6p$^1$) & 3.66\\ 
        NaZrCuSe$_3$ & Na(3s$^1$) Zr(4d$^2$5s$^2$) Cu(4s$^1$) & 3.67\\ 
        HfTlCuSe$_3$ & Hf(5d$^2$6s$^2$) Tl(6p$^1$) Cu(4s$^1$) & 3.69\\
        BaZrS$_3$ & Ba(6s$^2$) Zr(4d$^2$5s$^2$) & 3.60\\
        ZrTlCuSe$_3$ & Zr(4d$^2$5s$^2$) Tl(6p$^1$) Cu(4s$^1$) & 3.72\\
        \hline
    \end{tabular}
\end{table}

\begin{figure}[!htb]
\includegraphics[width=\textwidth]{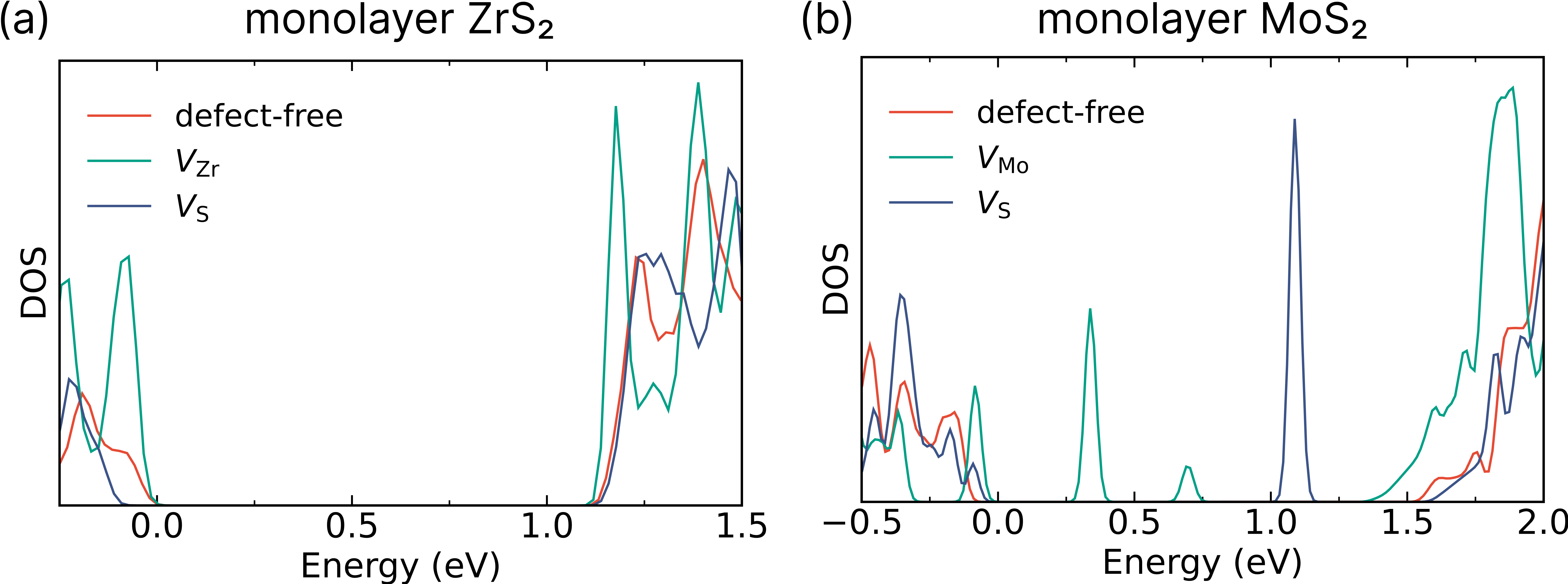}
\caption{\textbf{Defect tolerance of ZrS$_2$ compared to MoS$_2$.}
(a) Electronic DOS analysis for defect-free monolayer ZrS$_2$, monolayer ZrS$_2$ with Zr vacancy ($V_{Zr}$), and monolayer ZrS$_2$ with S vacancy ($V_{S}$).
The absence of defect states within the band gap confirms the defect-tolerant nature of the material.
(b) Electronic DOS analysis for defect-free monolayer MoS$_2$, monolayer MoS$_2$ with Mo vacancy ($V_{Mo}$), and monolayer MoS$_2$ with S vacancy ($V_{S}$).
The presence of prominent defect states deep within the band gap indicates the defect-intolerant nature of the material.}
\label{fig:defect_ZrS2}
\end{figure}

Semiconductors with an antibonding VBM and a bonding CBM have been predicted to be strongly tolerant of defects for both electrons and holes~\cite{zakutayev2014defect}, especially for optoelectronic applications. Intuitively, defect levels associated with unbonded cations or anions due to vacancies or interstitials should align with the orbital energies prior to hybridization and thus reside inside the bands rather than deep in the bandgap. In contrast, in semiconductors with a bonding VBM and an antibonding CBM, the orbital energies prior to hybridization lies deep within the bandgap, indicating that vacancies and interstitials can lead to deep in-gap states, as nicely illustrated in~\cite{zakutayev2014defect}. These deep in-gap states can act as effective recombination centers and are thus detrimental to applications that require long photocarrier lifetimes, such as photovoltaics. To verify the predicted defect tolerance in anti-gap semiconductors, we computed the defect energy levels in monolayer ZrS$_2$ caused by Zr or S vacancies. As shown in Fig.~\ref{fig:defect_ZrS2}a, both Zr and S vacancies cause resonances within the bands near the band edges in the total electronic DOS of a $6\times6\times1$ supercell of monolayer ZrS$_2$, confirming the absence of deep defect states in the gap. 
In contrast, the defect energy levels introduced by Mo or S vacancies in a $6\times6\times1$ supercell of monolayer MoS$_2$ reside deep within the band gap, as shown in Fig.~\ref{fig:defect_ZrS2}b. These deep-level states can act as trap centers, which can significantly reduce the efficiency of optoelectronic devices.

\begin{figure}[!htb]
\includegraphics[width=\textwidth]{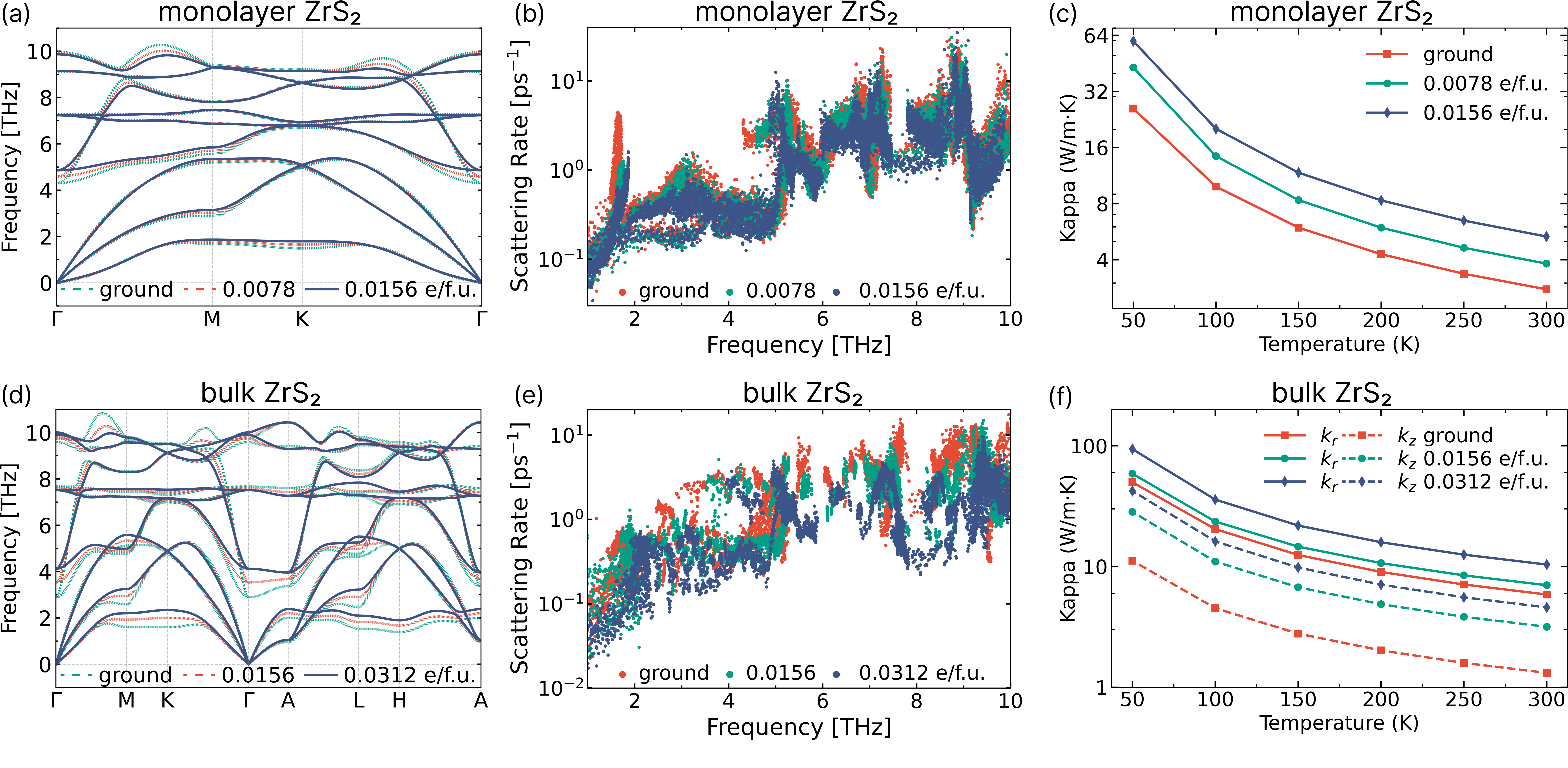}
\caption{\textbf{Calculated phonon and thermal transport properties of ZrS$_2$ with different photocarrier concentrations.}
(a) Phonon dispersion of monolayer ZrS$_2$ in the ground state ($n_{ph}=$ 0) and different excited states ($n_{ph}=$ 0.0078, 0.0156).
Increased phonon frequency is observed in acoustic modes and low-frequency optical modes with photoexcitation.
(b) Phonon-phonon scattering rate at 300 K in monolayer ZrS$_2$ in the ground state ($n_{ph}=$ 0) and different excited states ($n_{ph}=$ 0.0078, 0.0156).
The scattering rate decreases with photoexcitation.
(c) Temperature-dependent in-plane thermal conductivity ($\kappa_r$) in monolayer ZrS$_2$ in the ground state ($n_{ph}=$ 0) and different excited states ($n_{ph}=$ 0.0078, 0.0156).
The thermal conductivity exhibits a significant increase following photoexcitation.
(d) Phonon dispersion of bulk ZrS$_2$ in the ground state ($n_{ph}=$ 0) and different excited states ($n_{ph}=$ 0.0156, 0.0312), showing a similar stiffening effect as in the monolayer.
(e) Phonon-phonon scattering rate at 300 K in bulk ZrS$_2$ in the ground state ($n_{ph}=$ 0) and different excited states ($n_{ph}=$ 0.0156, 0.0312).
The scattering rate decreases with photoexcitation.
(f) Temperature-dependent thermal conductivity along in-plane ($\kappa_r$) and out-of-plane ($\kappa_z$) directions in bulk ZrS$_2$ in the ground state ($n_{ph}=$ 0) and different excited states ($n_{ph}=$ 0.0156, 0.0312).
Both the in-plane and out-of-plane thermal conductivities exhibit a significant increase with photoexcitation.
} 
\label{fig:phonon_ZrS2}
\end{figure}

Another unusual property of anti-gap semiconductors is photoinduced stiffening of the lattice. Intuitively, photoexcitation in an anti-gap semiconductor moves electrons from the antibonding VBM to occupy bonding CBM states. This transient electronic configuration will then drive the lattice to contract to minimize the total electronic energy. Here, we combine the density functional perturbation theory (DFPT) and the $\Delta$SCF method~\cite{martin2020electronic} to simulate this effect in ZrS$_2$. Recently, this method has been developed to directly evaluate the lattice properties in the presence of photoexcited charge carriers, enabling accurate modeling of photoinduced phonon softening and phase transitions in conventional semiconductors~\cite{lin2017ultrafast,krishnamoorthy2018semiconductor,peng2020topological,marini2021lattice}. More details of the method and the computational parameters can be found in the Supplementary Material. For both bulk and monolayer ZrS$_2$, we simulate how their lattice constants and bulk moduli change as a function of photocarrier concentration. As shown in Fig. S4, we observe a negative photoinduced strain and an increase of the bulk modulus with an increasing photocarrier concentration. This behavior is in sharp contrast to photoinduced destabilization in conventional semiconductors such as Si and GaAs. Then we further evaluate how photoexcitation affects phonon properties and thermal transport in both bulk and monolayer ZrS$_2$ with different photoexcited electrons per formula unit $n_{ph}$ (\textit{e}/f.u.) in Fig.~\ref{fig:phonon_ZrS2} (0.0156 \textit{e}/f.u. in monolayer ZrS$_2$ corresponds to a photocarrier concentration of $1.32\times10^{13}$ cm$^{-2}$ and 0.0312 \textit{e}/f.u. in bulk ZrS$_2$ corresponds to $4.55\times10^{20}$ cm$^{-3}$).
As depicted in Fig.~\ref{fig:phonon_ZrS2}a and ~\ref{fig:phonon_ZrS2}d, the phonon dispersions in monolayer and bulk ZrS$_2$ are changed in a similar fashion by photoexcitation.
In the presence of photocarriers, all three acoustic phonon branches along with the low-frequency optical phonon branches show increased frequencies with rising carrier concentrations, while the high-frequency optical branches remain nearly uninfluenced.
This phenomenon demonstrates the transient stabilization effect when photoexcitation induces a transition of electrons from antibonding to bonding states and suggests that anti-gap semiconductors generally show a negative and strong photostriction~\cite{xiang2024high}. 
Consequently, the strengthened interatomic bonding resulting from photoinduced phonon stiffening suppresses anharmonic phonon-phonon interactions, leading to a reduced phonon-phonon scattering rate, as illustrated in Fig.~\ref{fig:phonon_ZrS2}b and ~\ref{fig:phonon_ZrS2}e.
To further investigate the influence of photoinduced phonon stiffening on thermal transport properties, Fig.~\ref{fig:phonon_ZrS2}c and \ref{fig:phonon_ZrS2}f present the calculated lattice thermal conductivity of monolayer and bulk ZrS$_2$ with different photocarrier concentrations from 50 K to 300 K.
For monolayer ZrS$_2$, the in-plane thermal conductivity ($\kappa_r$) increases from 2.8 W/(m K) in the dark state to 5.3 W/(m K) with $n_{ph}$ = 0.0156 (\textit{e}/f.u.) at 300 K, marking a 90\% enhancement.
For bulk ZrS$_2$,  the out-of-plane thermal conductivity ($\kappa_z$) undergoes a remarkable 253\% enhancement, rising from 1.3 W/(m·K) to 4.6 W/(m·K) as $n_{ph}$ increases from 0 to 0.0312 (\textit{e}/f.u.) at 300 K.
Meanwhile, the in-plane thermal conductivity ($\kappa_r$) exhibits a 76\% increase, elevating from 5.9 W/(m·K) to 10.4 W/(m·K) under the same conditions.
We note that these results only provide a lower bound on the thermal conductivity enhancement in anti-gap semiconductors under photoexcitation, since the contribution of bipolar thermal transport from photoexcited electron-hole pairs can further enhance the thermal conductivity~\cite{drabble1961thermal}.

% Subheadings can be used in this section to organize the data/discussion. First describe the results as presented in the Figure. Then provide a discussion of the result (explain why the result makes sense, compare to previous results, discuss the importance/indications of the result, how does the result contribute to the main objective of this paper).

% \section{Conclusion}
% Briefly summarize the approach and the main results. Provide a more detailed discussion of the significance/impact of this work. If applicable, briefly provide an outlook for related future work.

% In summary, we applied ... to study.... We found that.... Our work opens an avenue to.../provides guidelines for.... Future work. 
In summary, our study explores a new class of semiconductors, termed anti-gap semiconductors, characterized by an anti-bonding VBM and a bonding CBM. Using high-throughput computational screening, we identify over 100 anti-gap semiconductors and analyze in detail the TMD family MX$_2$ (M=Hf, Zr; X=S, Se), confirming their defect tolerance through first-principles calculations. A key discovery is that photoexcitation in these materials leads to lattice stiffening and an increase in thermal conductivity, contrary to the usual lattice softening observed in conventional semiconductors. This effect, driven by electron transitions from anti-bonding to bonding states, suggests potential applications of these materials for photo-driven thermal conductivity switching. The findings provide a foundation for further research into anti-gap semiconductors and their unique photophysical properties.

\begin{acknowledgments}
This work is based on research supported by the U.S. Office of Naval Research under award number N00014-22-1-2262. This work used Stampede2 at Texas Advanced Computing Center (TACC) through allocation MAT200011 from the Advanced Cyberinfrastructure Coordination Ecosystem: Services \& Support (ACCESS) program, which is supported by National Science Foundation grants 2138259, 2138286, 2138307, 2137603, and 2138296. Use was also made of computational facilities purchased with funds from the National Science Foundation (award number CNS-1725797) and administered by the Center for Scientific Computing (CSC) at University of California, Santa Barbara (UCSB). The CSC is supported by the California NanoSystems Institute and the Materials Research Science and Engineering Center (MRSEC; NSF DMR-2308708) at UCSB.
\end{acknowledgments}

\clearpage

\bibliography{references.bib}% Produces the bibliography via BibTeX.

\end{document}

%\preprint{APS/123-QED}

\title{Supplementary Material: High-throughput Discovery of Anti-gap Semiconductors} 

\author{Zeyu Xiang}
\affiliation{Department of Mechanical Engineering, University of California, Santa Barbara, CA 93106, USA}

\author{Fanghao Zhang}
\affiliation{Department of Mechanical Engineering, University of California, Santa Barbara, CA 93106, USA}

\author{Bolin Liao}
\email{bliao@ucsb.edu} \affiliation{Department of Mechanical Engineering, University of California, Santa Barbara, CA 93106, USA}

\maketitle

%\tableofcontents
%\renewcommand\linenumberfont{\normalfont\tiny}

%\linenumbers\relax % Commence numbering lines

\section{Computational Methods}
Density functional theory (DFT) calculations were accomplished
via the Vienna ab initio simulation package (VASP)~\cite{kresse1996efficiency,kresse1996efficient} based on
the projected augmented wave pseudopotentials~\cite{blochl1994projector}. 
The Perdew-Burke-Ernzerhof form of generalized gradient approximation (PBE-GGA)~\cite{perdew1996generalized} was used for structural optimization.
For the bulk structure of MX$_2$ (M=Hf, Zr; X=S, Se), plane-wave cutoff energy convergence and \textit{k}-point convergence  were firstly tested to ensure the lattice parameters were well relaxed, yielding optimal values of 500 eV and $\Gamma$-centered $12\times12\times6$ grid, respectively.
The energy and force convergence criteria were set to $1\times10^{-7}$ eV and $1\times10^{-3}$ eV/\AA, respectively.
Besides, Grimmes's semiempirical van der Waals corrections~\cite{grimme2006semiempirical} were considered for the bulk forms due to the layered structures.  
For the monolayer calculation, a vacuum region of 20 \AA\ was applied to separate the layers.

For high-throughput search of anti-gap semiconductors, nearly 15,000 stable materials with band gaps below 5 eV (calculated by DFT-PBE) were screened from the Materials Project database~\cite{jain2013commentary}.
A plane-wave cutoff energy of 400 eV was first utilized for the lattice relaxation, followed by the COHP calculations performed using the Local Orbital Basis Suite Towards Electronic-Structure Reconstruction (LOBSTER) software~\cite{maintz2016lobster}. 
COHP analysis decomposes the energy of the electronic band structure into interactions (overlaps) between pairs of atomic orbitals between adjacent atoms.
In other words, it provides a bond-weighted electronic density of states, allowing for a quantitative measure of the bonding and antibonding contributions to the electronic bands, especially near the band edges.
By convention, a positive (negative) sign indicates antibonding (bonding) interactions and COHP diagrams plot the negative value (-pCOHP), therefore making bonding (antibonding) states on the right (left) of the axis for intuitive visualization~\cite{deringer2011crystal}.
PbeVaspFit2015~\cite{maintz2016lobster} is selected for reconstructing chemical bonding information from DFT calculations.
The energy window for LOBSTER calculations is set from -35 eV to 5 eV, with a Gaussian smearing width of 0.01 eV and an atomic pair distance range of 0.1 to 6 \AA. For the defect-tolerance verification calculation, a large $6\times6\times1$ ZrS$_2$ supercell with 108 atoms in combination with $2\times2\times1$ \textit{k}-point grids was adopted after convergence test.

The phonon dispersion relations were determined through finite-displacement force calculations, from which the harmonic interatomic force constants (IFCs) were extracted~\cite{broido2007intrinsic}.
The dynamic matrices were then constructed and diagonalized using the Phonopy package to obtain the phonon eigenfrequencies~\cite{togo2015first,togo2023first}.
The convergence of the phonon dispersion with respect to the supercell size and \textit{k}-point grid was thoroughly tested.
We employed $4\times4\times2$ and $4\times4\times1$ supercells to calculate the phonon properties of bulk and monolayer forms, respectively.
The supercell \textit{k}-point grids are converged as $2\times2\times1$ for monolayers, $1\times1\times1$ for bulk ZrS$_2$ and HfS$_2$, and $2\times2\times1$ for bulk ZrSe$_2$ and HfSe$_2$.
The anharmonic third-order IFCs were computed with the identical supercell size and \textit{k}-mesh through the finite-displacement method, and the cutoff for neighboring interactions was fully tested to ensure convergence.
The lattice thermal conductivity $k_{ph}$ was then determined by iteratively solving the phonon Boltzmann transport equation (BTE) using the ShengBTE package~\cite{li2014shengbte}, following a convergence test of the \textit{q}-mesh sampling in phonon momentum space.
The $\Delta$-SCF method was used to simulate materials with a fixed concentration of photoexcited electrons, where the occupations of the electronic state near the band edge were fixed during the simulation~\cite{paillard2017ab}.

\clearpage

\section{Additional Data and Analysis}

\subsection{COHP analysis for monolayer and bulk structures of MX$_2$ (M=Hf, Zr; X=S, Se) family}

\begin{figure}[!htb]
\includegraphics[width=\textwidth]{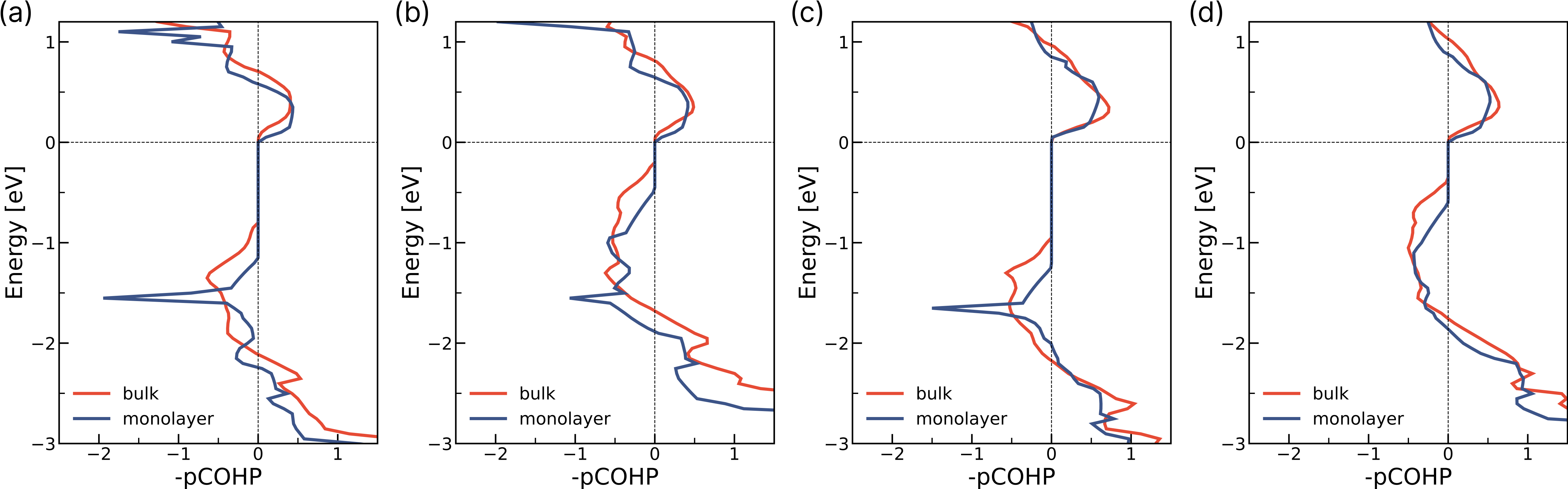}
\caption{
(a) COHP diagram of monolayer and bulk ZrS$_2$, showing the anti-bonding valence band and bonding conduction band.
(b) COHP diagram of monolayer and bulk ZrSe$_2$, showing the anti-bonding valence band and bonding conduction band.
(c) COHP diagram of monolayer and bulk HfS$_2$, showing the anti-bonding valence band and bonding conduction band.
(d) COHP diagram of monolayer and bulk HfSe$_2$, showing the anti-bonding valence band and bonding conduction band.
} 
\label{fig:figCu3N}
\end{figure}

\clearpage

\subsection{COHP analysis for Cu$_3$N}
Cu$_3$N was reported to be defect-tolerant considering its antibonding valence band and bonding conduction band in previous literature~\cite{zakutayev2014defect}.
However, we find that its conduction band  is actually antibonding instead of bonding after a detailed analysis based on COHP.
The lattice was relaxed with the Heyd-Secuseria-Ernzerhof (HSE06) exchange-correlation functional~\cite{heyd2003hybrid}, with the lattice parameter of 3.80 \AA\ and the bandgap of 0.90 eV ($R\xrightarrow{}M$).

\begin{figure}[!htb]
\includegraphics[width=0.75\textwidth]{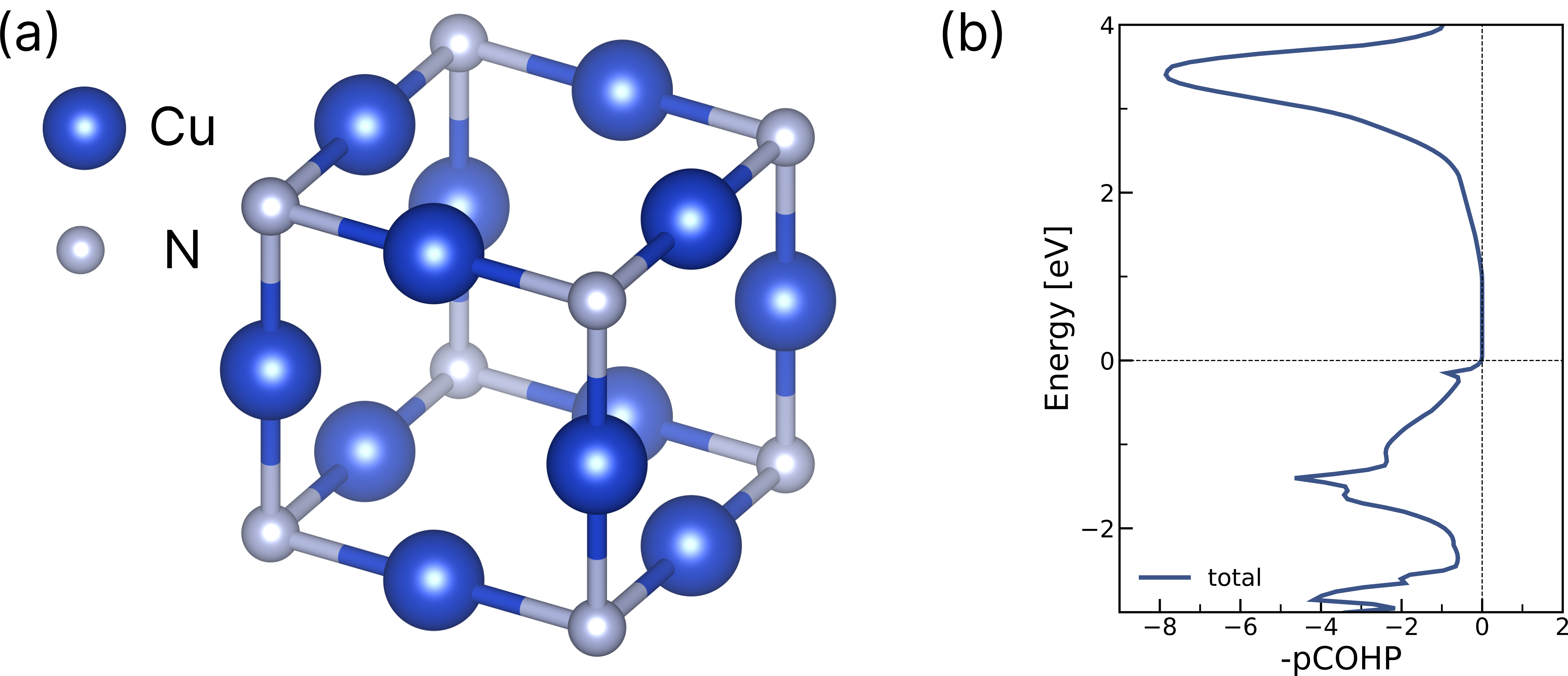}
\caption{
(a) Crystal structure of cubic Cu$_3$N that is the same with the previous literature~\cite{zakutayev2014defect}.
(b) COHP diagram of Cu$_3$N after lattice relaxation with HSE06 exchange-correlation functional. 
It reveals that both VBM and CBM exhibit antibonding characteristics.
} 
\label{fig:figCu3N}
\end{figure}

\clearpage

\subsection{COHP analysis and schematic diagrams of orbital interactions for MoS$_2$}

\begin{figure}[!htb]
\includegraphics[width=0.75\textwidth]{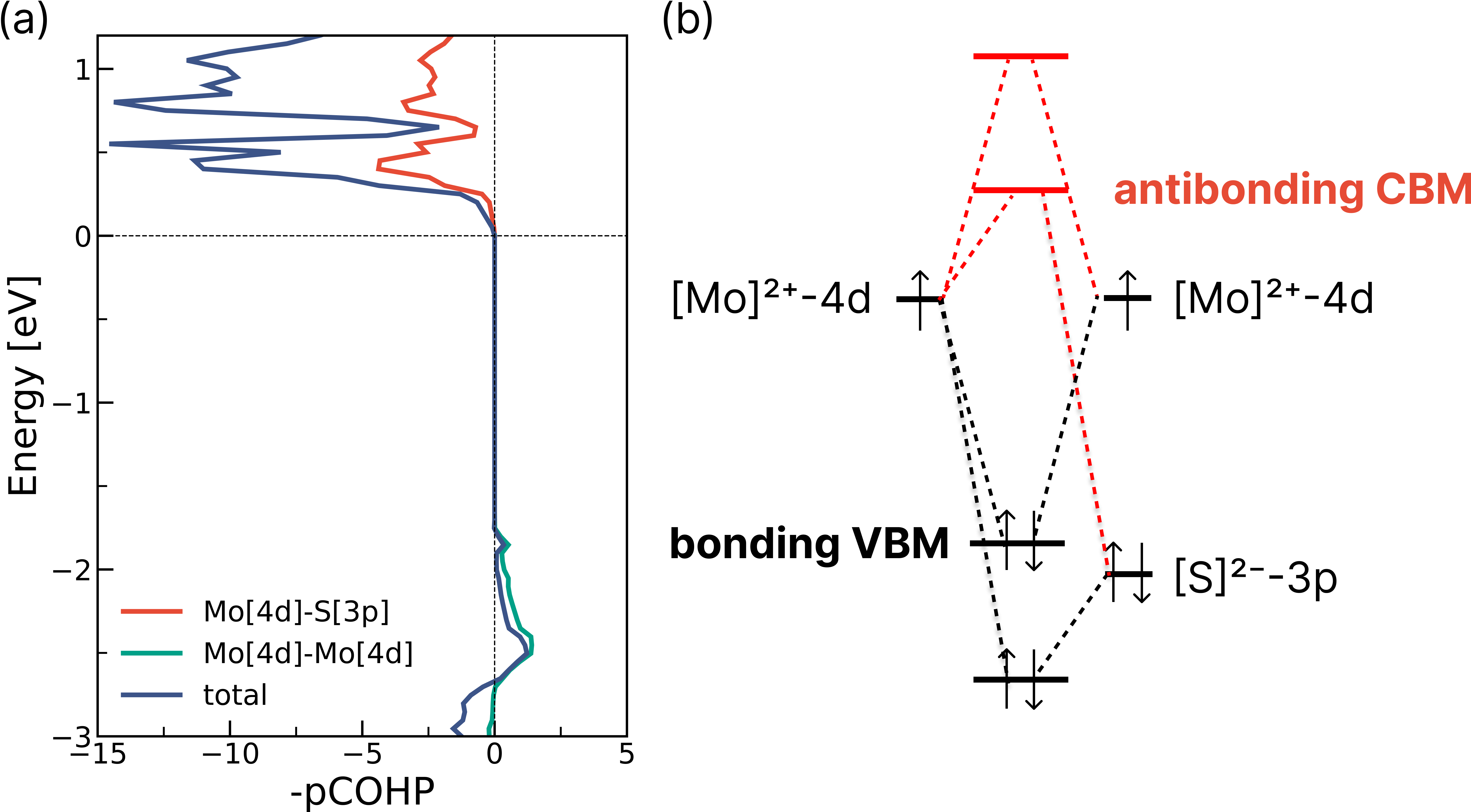}
\caption{
(a) COHP diagram of MoS$_2$, showing the bonding VBM and antibonding CBM.
(b) The orbital interaction diagram shows that Mo-4d and S-3p states hybridize to form the antibonding CBM and Mo-4d and Mo-4d states hybridize to form the bonding VBM.
} 
\label{fig:figMoS2}
\end{figure}

\clearpage

\subsection{Photoinduced strain and bulk modulus variations with photocarrier concentration for MX$_2$ (M=Hf, Zr; X=S, Se) family}
The negative photoinduced strain ($\epsilon$), coupled with the increase in the bulk modulus ($B=-\frac{dP}{dlnA}$) as the photocarrier concentration ($n_{ph}$) increases, aligns well with the anti-gap characteristic observed in the monolayer MX$_2$ (M=Hf, Zr; X=S, Se) family.

\begin{figure}[!htb]
\includegraphics[width=\textwidth]{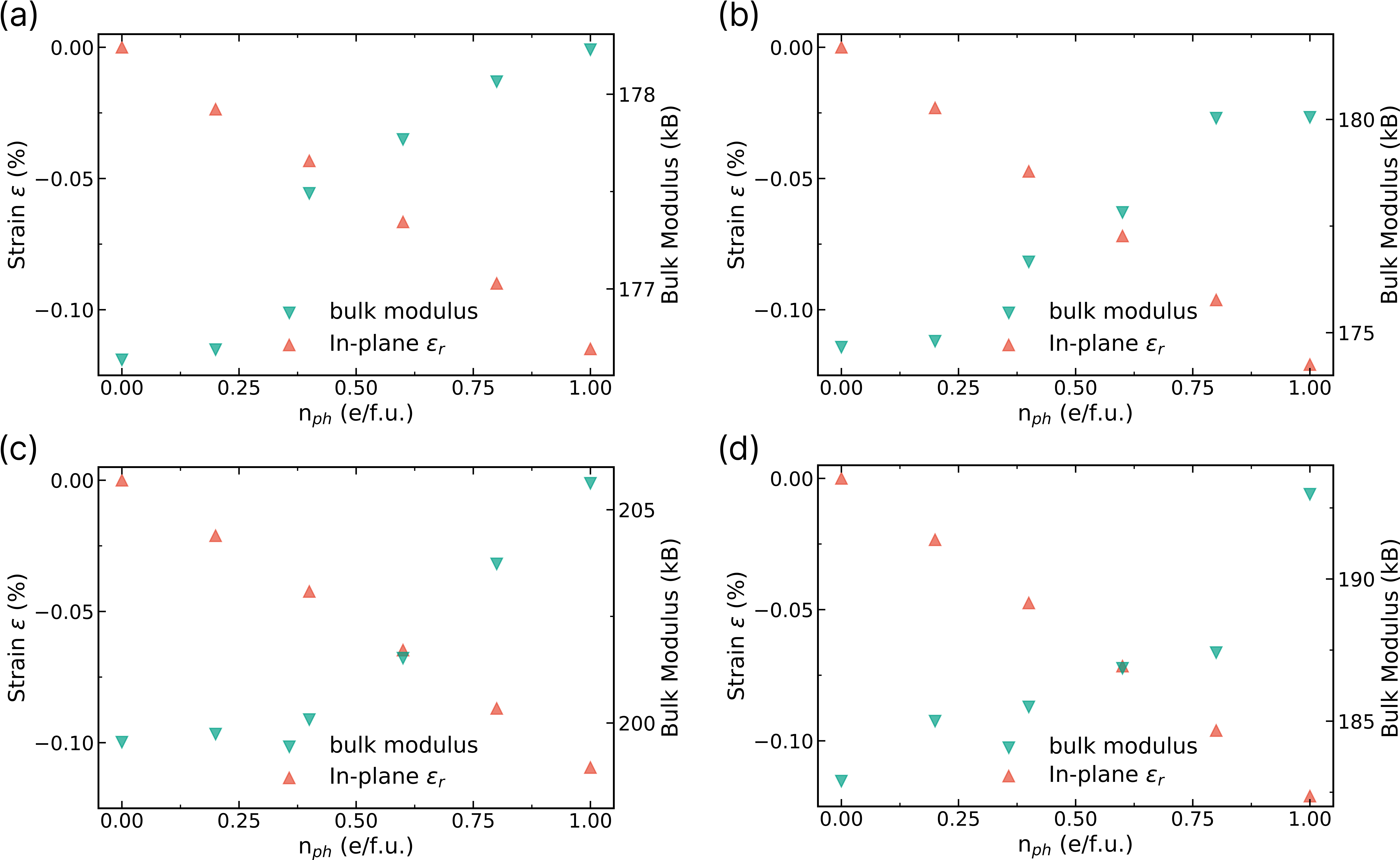}
\caption{
Photostriction and bulk modulus changes with photoexcited carrier concentrations ($n_{ph}=$ 0, 0.2, 0.4, 0.6, 0.8, 1.0 \textit{e}/f.u.) of
(a) monolayer ZrS$_2$,
(b) monolayer ZrSe$_2$,
(c) monolayer HfS$_2$,
and (d) monolayer HfSe$_2$.
} 
\label{fig:figCu3N}
\end{figure}

\clearpage

\subsection{Curvature effective mass calculations of electrons (CBM) and holes (VBM) for both bulk and monolayer structure of MX$_2$ (M=Hf, Zr; X=S, Se) family}
Generally, the effective mass refers to the curvature effective mass ($m_c$), which is defined as:
\begin{equation}
    m_c=\frac{1}{\hbar^2}\frac{\partial^2E}{\partial k^2}.
    \label{effmass}
\end{equation}
Here we have listed the effective mass for electrons and holes along different directions in reciprocal space in the following tables.

\begin{table}[h] 
    \centering 
    \renewcommand{\arraystretch}{1.2}
    \caption{Curvature effective mass of electrons and holes in bulk MX$_2$ (M=Hf, Zr; X=S, Se).}
    \label{tab:effmass_bulk}  
    \begin{tabular}{p{2cm}p{2cm}p{2cm}p{2cm}p{2cm}p{2cm}p{2cm}}
        \hline
        & \multicolumn{3}{c}{hole effective mass} & \multicolumn{3}{c}{electron effective mass} \\    % Table header
        \hline
        & [100] & [110] & [001] & [100] & [1$\overline{2}$0] & [001] \\ 
        \hline
        ZrS$_2$ & 0.44 & 0.44 & 13.50 & 1.70 & 0.30 & 2.37 \\  
        ZrSe$_2$ & 0.30 & 0.29 & 6.77 & 1.46 & 0.21 & 1.69 \\ 
        HfS$_2$ & 0.47 & 0.46 & 14.97 & 2.12 & 0.25 & 2.66 \\ 
        HfSe$_2$ & 0.34 & 0.34 & 9.52 & 1.94 & 0.19 & 2.18 \\ 
        \hline
    \end{tabular}
\end{table}

\begin{table}[h] 
    \centering 
    \renewcommand{\arraystretch}{1.2}
    \caption{Curvature effective mass of electrons and holes in monolayer MX$_2$ (M=Hf, Zr; X=S, Se).}
    \label{tab:effmass_monolayer}  
    \begin{tabular}{p{2cm}p{2cm}p{2cm}p{2cm}p{2cm}}
        \hline
        & \multicolumn{2}{c}{hole effective mass} & \multicolumn{2}{c}{electron effective mass} \\    % Table header
        \hline
        & [100] & [110] & [100] & [1$\overline{2}$0] \\ 
        \hline
        ZrS$_2$ & 0.46 & 0.45 & 2.14 & 0.30 \\  
        ZrSe$_2$ & 0.32 & 0.32 & 2.08 & 0.22 \\  
        HfS$_2$ & 0.49 & 0.48 & 2.51 & 0.25 \\  
        HfSe$_2$ & 0.36 & 0.35 & 2.50 & 0.19 \\  
        \hline
    \end{tabular}
\end{table}
\clearpage

\subsection{Atomic configuration for monolayer ZrS$_2$ and MoS$_2$ with vacancies}

\begin{figure}[!htb]
\includegraphics[width=0.75\textwidth]{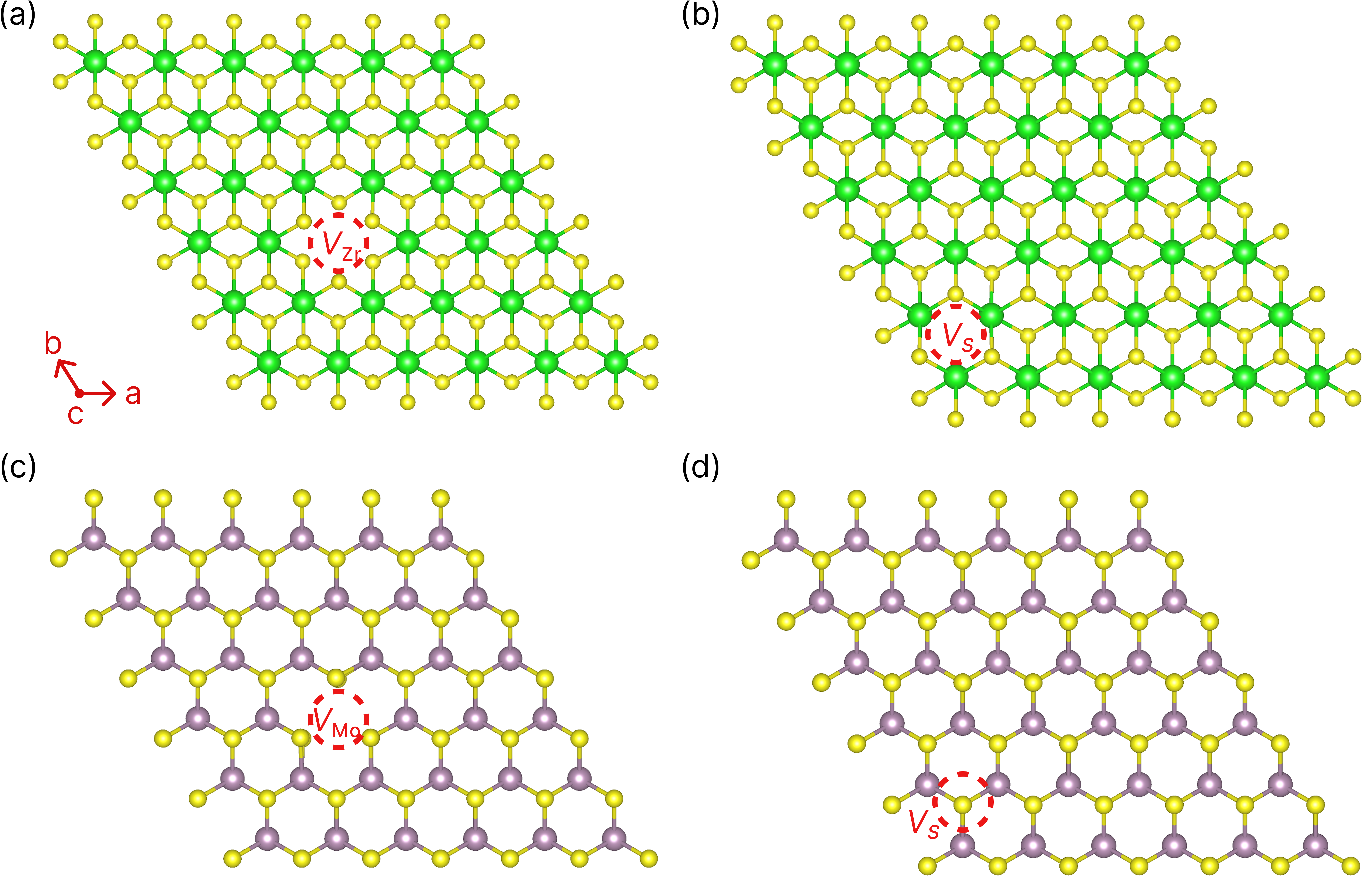}
\caption{
(a) Atomic configuration of monolayer ZrS$_2$ with Zr vacancy ($V_\mathrm{Zr}$).
(b) Atomic configuration of monolayer ZrS$_2$ with S vacancy ($V_\mathrm{S}$).
(c) Atomic configuration of monolayer MoS$_2$ with Mo vacancy ($V_\mathrm{Mo}$).
(d) Atomic configuration of monolayer MoS$_2$ with S vacancy ($V_\mathrm{S}$).
} 
\label{fig:figCu3N}
\end{figure}
\clearpage

\subsection{Candidate semiconductors with antibonding valence bands and bonding conduction bands}
The following table contains the screened stable semiconductors without rare earth elements from the Materials Project~\cite{jain2013commentary} with both antibonding valence bands and bonding conduction bands.
The strength of the bonding feature of conduction bands and the antibonding feature of valence bands are quantified by integrating the area under the COHP curve within 0.1 eV above CBM and below VBM, respectively.
The list was sorted by the strength of the bonding feature within 0.1 eV above the CBM. The Materials Project ID number is provided after each material name. 

\setlength{\tabcolsep}{12pt}
\begin{longtable}{p{6cm} p{4cm} p{4cm} }
    \caption{List of MX$_2$ (M=Hf, Zr; X=S, Se) family. Second and third column contain integrated COHP within 0.1 eV above CBM and below VBM.} \\
         \hline
         \ MX$_2$ (M=Hf, Zr; X=S, Se) & \thead{0.1 eV above CBM}   & \thead{0.1 eV below VBM}  \\
         \hline
         ZrS$_2$ & 2.7961 & -0.7716 \\
         ZrSe$_2$ & 1.5746 & -2.1136 \\
         HfS$_2$ & 1.7132 & -1.3183 \\
         HfSe$_2$ & 3.8201 & -1.0996 \\
         \hline
\end{longtable}

\setlength{\tabcolsep}{12pt}
\begin{longtable}{p{6cm} p{4cm} p{4cm} }
    \caption{List of anti-gap semiconductors. Second and third column contain integrated COHP within 0.1 eV above CBM and below VBM. Bandgap is calculated with DFT-PBE and likely underestimated.} \\
         \hline
         \ $E_g<1$ eV  & \thead{0.1 eV above CBM}   & \thead{0.1 eV below VBM}  \\
         \hline
         \endfirsthead
         \hline
         \ Material & \thead{0.1 eV below VBM}   & \thead{0.1 eV above CBM}  \\
         \hline
         \endhead
         \hline
         \endfoot
         Tl$_2$O-mp$27484$ & 11.6699 & -0.3933 \\
         Th$_2$SbNO-mp$1217259$ & 10.4241 & -1.5436 \\
         Sr$_2$Li$_7$CuN$_4$-mp$1218762$ & 6.0890 & -1.7086 \\
         RbUO$_3$-mp$11775$ & 5.4668 & -0.9173 \\
         ZrTl$_2$S$_3$-mp$1102830$ & 5.1008 & -0.0189 \\
         Hf$_3$Tl$_2$(CuSe$_4$)$_2$-mp$570700$ & 4.5739 & -5.6916 \\
         Zr$_3$Tl$_2$(CuS$_4$)$_2$-mp$1207410$ & 4.5399 & -1.8754 \\
         Zr$_3$Tl$_2$(CuSe$_4$)$_2$-mp$1207399$ & 4.1104 & -0.0279 \\
         NaHfCuSe$_3$-mp$505448$ & 4.0548 & -0.0991 \\
         HfTl$_2$Se$_3$-mp$1102545$ & 3.9759 & -2.1410 \\
         NaZrCuSe$_3$-mp$505172$ & 3.9365 & -0.6038 \\
         HfSe$_2$-mp$985831$ & 3.8201 & -1.0996 \\
         HfTlCuSe$_3$-mp$9397$ & 3.7797 & -1.8478 \\
         BaZrS$_3$-mp$540771$ & 3.5335 & -1.8655 \\
         ThSeO-mp$7950$ & 3.4530 & -3.2712 \\
         Th(CuP)$_2$-mp$9581$ & 3.3869 & -0.0317 \\
         Zr$_2$SeN$_2$-mp$1079726$ & 3.0712 & -0.4804 \\
         ZrTlCuSe$_3$-mp$7050$ & 2.9307 & -0.5825 \\
         NaZrCuS$_3$-mp$9107$ & 2.9092 & -4.0476 \\
         ZrSnS$_3$-mp$17324$ & 2.7068 & -6.7427 \\
         HfIN-mp$567441$ & 2.5046 & -8.5500 \\
         Ba$_3$Zr$_2$S$_7$-mp$8570$ & 2.4981 & -19.6101 \\
         LiAcRh$_2$-mp$1185329$ & 2.3362 & -2.1870 \\
         ThTeO-mp$3718$ & 2.2874 & -4.2092 \\
         AcH$_3$-mp$861605$ & 2.0853 & -0.2756 \\
         NaHf$_2$CuSe$_5$-mp$571189$ & 2.0635 & -10.2674 \\
         NaTiCuS$_3$-mp$505171$ & 1.9462 & -6.2648 \\
         NbTlO$_3$-mp$757636$ & 1.7174 & -11.0663 \\
         ZrSe$_2$-mp$2076$ & 1.5746 & -2.1136 \\
         ZrTlCuS$_3$-mp$7049$ & 1.5636 & -4.1681 \\
         TaTlO$_3$-mp$676262$ & 1.5226 & -1.2660 \\
         TlInS$_2$-mp$20042$ & 1.5056 & -0.3620 \\
         Zr$_2$SN$_2$-mp$11583$ & 1.4289 & -0.2510 \\
         Hf$_2$SeN$_2$-mp$1029330$ & 1.1827 & -0.1510 \\
         RbHfAgTe$_3$-mp$9780$ & 0.9507 & -9.8786 \\
         BaZrN$_2$-mp$3104$ & 0.9324 & -0.4719 \\
         TiGeS$_3$-mp$1105715$ & 0.8449 & -5.1988 \\
         Hf$_2$SN$_2$-mp$1017567$ & 0.7568 & -0.1200 \\
         ThSe$_2$-mp$7951$ & 0.7396 & -1.7503 \\
         CaHfN$_2$-mp$1029359$ & 0.6226 & -0.5712 \\
         TiPN$_3$-mp$989624$ & 0.6187 & -0.2228 \\
         MgTiN$_2$-mp$1245745$ & 0.6116 & -1.0168 \\
         SrTiN$_2$-mp$9517$ & 0.4837 & -1.0331 \\
         CaZrN$_2$-mp$1029267$ & 0.3933 & -0.6188 \\
         NaAcTe$_2$-mp$865081$ & 0.3574 & -2.2437 \\
         LiAcTe$_2$-mp$864755$ & 0.3376 & 0.0000 \\
         TiBrN-mp$27849$ & 0.3308 & -0.4071 \\
         TiS$_3$-mp$9920$ & 0.2666 & -0.3456 \\
         Na$_2$Zr(CuS$_2$)$_2$-mp$556536$ & 0.2480 & -6.2163 \\
         TiNCl-mp$27850$ & 0.1996 & -0.4364 \\
         Ca$_3$BiN-mp$31149$ & 0.1338 & -0.7004 \\
         Ca$_3$PN-mp$11824$ & 0.0889 & -0.7113 \\
         Ca$_3$SbN-mp$1013548$ & 0.0471 & -0.7104 \\
         RbVN$_2$-mp$1029312$ & 0.0466 & -0.3295 \\
         Li$_3$H$_4$Rh-mp$697047$ & 0.0439 & -2.6162 \\
         \hline
         \ $1<E_g<2$ eV & \thead{0.1 eV below VBM}   & \thead{0.1 eV above CBM}  \\
         \hline
         ThTlCuSe$_3$-mp$1102015$ & 25.6551 & -15.6777 \\
         Tl$_5$SbO$_5$-mp$27220$ & 7.4240 & -5.5424 \\
         UTl$_2$O$_4$-mp$1025154$ & 7.3127 & -2.8708 \\
         Cs$_2$U$_2$O$_7$-mp$548479$ & 6.7084 & -0.5688 \\
         HfSnS$_3$-mp$8725$ & 6.1403 & -6.4897 \\
         K$_2$U$_2$O$_7$-mp$29361$ & 5.1856 & -10.3622 \\
         Na$_5$NpO$_6$-mp$1095270$ & 5.1785 & 0.0000 \\
         MgUO$_4$-mp$2271432$ & 4.7806 & -0.1555 \\
         ThS$_2$-mp$1146$ & 4.5860 & -1.7513 \\
         UPbO$_4$-mp$504922$ & 4.3693 & -6.3326 \\
         Rb$_2$U$_2$O$_7$-mp$560458$ & 3.9995 & -29.4801 \\
         Tl$_6$TeO$_6$-mp$8387$ & 3.9886 & -15.2180 \\
         ThSO-mp$8136$ & 3.9816 & -0.0022 \\
         ZrS$_2$-mp$1186$ & 2.7961 & -0.7716 \\
         ZrBrN-mp$541912$ & 1.9754 & -0.0013 \\
         WO$_3$-mp$2383161$ & 1.9672 & 0.0000 \\
         Na$_2$UO$_4$-mp$554191$ & 1.9255 & -14.7596 \\
         ZrS$_3$-mp$9921$ & 1.7357 & -1.9675 \\
         HfMgN$_2$-mp$1029834$ & 1.7307 & -0.1111 \\
         HfS$_2$-mp$985829$ & 1.7132 & -1.3183 \\
         Rb$_3$NpH$_8$O$_9$-mp$1198189$ & 1.3316 & -248.9383 \\
         BaHfS$_3$-mp$1105549$ & 0.9299 & -9.7258 \\
         UTl$_2$(TeO$_4$)$_2$-mp$560464$ & 0.8604 & -33.9548 \\
         SrTh$_2$Se$_5$-mp$17282$ & 0.7520 & -18.5746 \\
         Tl$_3$BO$_3$-mp$4584$ & 0.5190 & -1.2101 \\
         UCdO$_4$-mp$545598$ & 0.4983 & -1.9432 \\
         NaTaN$_2$-mp$5475$ & 0.4970 & -2.4127 \\
         Sr$_3$Ti$_2$O$_7$-mp$3349$ & 0.4037 & -9.4565 \\
         Sr$_2$TiO$_4$-mp$5532$ & 0.4008 & -0.1453 \\
         Ta$_3$O$_7$F-mp$753747$ & 0.3398 & -0.7226 \\
         Sr$_4$Ti$_3$O$_{10}$-mp$31213$ & 0.1140 & -8.4440 \\
         \hline
         \ $2<E_g<5$ eV & \thead{0.1 eV below VBM}   & \thead{0.1 eV above CBM}  \\
         \hline
         KThCuS$_3$-mp$12365$ & 20.1876 & -0.0479 \\
         Li$_2$UO$_4$-mp$9607$ & 10.1347 & -136.6552 \\
         AcI$_3$-mp$861867$ & 7.7890 & -0.6327 \\
         Na$_4$UO$_5$-mp$5327$ & 6.9436 & -4.3746 \\
         Li$_4$UO$_5$-mp$7714$ & 5.8161 & -0.0633 \\
         CaU$_2$(BO$_5$)$_2$-mp$1227453$ & 5.3619 & -108.6335 \\
         K$_4$CaU(Si$_2$O$_7$)$_2$-mp$1195727$ & 4.1879 & -47.6822 \\
         AcAlO$_3$-mp$1183115$ & 1.9238 & -1.8138 \\
         K$_4$SrU$_3$O$_12$-mp$22634$ & 1.8573 & -79.4322 \\
         K$_2$SrTa$_2$O$_7$-mp$7148$ & 1.5843 & -5.8299 \\
         TlF-mp$720$ & 1.3808 & -27.5703 \\
         PaTlO$_3$-mp$862832$ & 0.7256 & -2.8562 \\
         Ta$_2$O$_5$-mp$1539317$ & 0.6230 & -4.1686 \\
         NaTaO$_3$-mp$3858$ & 0.5169 & -3.3264 \\
         BaSrTa$_2$O$_7$-mp$559151$ & 0.5140 & -13.6851 \\
         CsSr$_2$Ta$_3$O$_10$-mp$7181$ & 0.4635 & -5.6474 \\
         Ba$_2$ZrO$_4$-mp$8335$ & 0.4440 & -0.0803 \\
         Ba$_3$Zr$_2$O$_7$-mp$755895$ & 0.3061 & -13.1216 \\
         Tl$_3$SiF$_7$-mp$1133284$ & 0.3032 & -12.6447 \\
         BaZrO$_3$-mp$1019544$ & 0.2821 & -0.9952 \\
         CaTiO$_3$-mp$4019$ & 0.2690 & -1.2530 \\
         K$_3$Ta$_3$(BO$_6$)$_2$-mp$9870$ & 0.2675 & -5.6169 \\
         TiTl$_2$O$_3$-mp$17986$ & 0.2201 & -14.7660 \\
         BaHfO$_3$-mp$998552$ & 0.1940 & -6.7187 \\
         KTaO$_3$-mp$3614$ & 0.1710 & -6.0763 \\
         Li$_2$HfO$_3$-mp$2724101$ & 0.1648 & -1.1055 \\
         Ta$_12$MoO$_33$-mp$1217800$ & 0.1365 & -3.8344 \\
         PbF$_2$-mp$315$ & 0.0520 & -14.9532 \\
         Ba$_4$Ta$_10$FeO$_30$-mp$1228547$ & 0.0442 & -24.4706 \\
         TiO$_2$-mp$390$ & 0.0441 & -0.2480 \\
         HfBrN-mp$568346$ & 0.0138 & -0.3862 \\
         \hline
\end{longtable}
\clearpage

\bibliography{references.bib}